\documentstyle[twoside,titlepage,cite,12pt]{article}
\begin{document}
\font\utap=cmmib10 scaled\magstep2
\font\urc=cmmib10 scaled\magstephalf
\font\smbf=cmbx10 scaled\magstep2
\font\smit=cmti12 
\font\fontname=cmcsc10 
\newcommand{\himpc}{{\hbox {$h^{-1}$}{\rm Mpc}} }
\newcommand{\gtsima}{$\; \buildrel > \over \sim \;$}
\newcommand{\ltsima}{$\; \buildrel < \over \sim \;$}
\newcommand{\simgt}{\lower.5ex\hbox{\gtsima}}
\newcommand{\simlt}{\lower.5ex\hbox{\ltsima}}
\newcommand{\avr}[1]{\left\langle #1 \right\rangle}
\newcommand{\cum}[1]{\left\langle #1 \right\rangle_{\rm c}}
\newcommand{\bfm}[1]{{\mbox{\boldmath $#1$}}}
\newcommand{\sbfm}[1]{{\mbox{\footnotesize\boldmath $#1$}}}
\newcommand{\bfq}{\bfm{q}}
\newcommand{\bfv}{\bfm{v}}
\newcommand{\bfx}{\bfm{x}}
\newcommand{\bfPsi}{\bfm{\Psi}}
\def\pp{\par\parshape 2 0truecm 15.5truecm 1truecm 14.5truecm\noindent}
\renewcommand{\theequation}{\mbox{\rm {\arabic{equation}}}}
\renewcommand{\thesection}{\normalsize\bf\arabic{section}.}
\renewcommand{\thesubsection}
{\normalsize\it\arabic{section}.\normalsize\it\arabic{subsection}.}
\renewcommand{\thepage}{-- \arabic{page} --}

\begin{titlepage}

\noindent {\utap UTAP}\hspace*{0.2cm}
{\urc the \hspace*{0.1cm} University \hspace*{0.1cm} of
\hspace*{0.1cm} Tokyo \hspace*{0.1cm} \hspace*{0.2cm} Theoretical
\hspace*{0.1cm} Astrophysics}

\vspace*{-0.6cm}
\hspace*{13.5cm}UTAP-210/95

\hspace*{13.5cm}October, 1995

\vspace*{1cm}

\addtocounter{footnote}{1}

\baselineskip=24pt

\begin{center}

{\large \bf  On Second-order Perturbation Theories of Gravitational
Instability in Friedmann-Lema\^{\i}tre Models}

\vspace{1.0cm}

{\sc Takahiko Matsubara}

\vspace{0.5cm}
\noindent {\it Department of Physics, The University of Tokyo,
Tokyo 113, Japan.}

\noindent { e-mail: matsu@utaphp1.phys.s.u-tokyo.ac.jp}

\vspace{0.6cm}

\end{center}
\vspace{1.0cm}
\centerline {\bf ABSTRACT}
\baselineskip=20pt

The Eulerian and Lagrangian second-order perturbation theories are
solved explicitly in closed forms in $\Omega \neq 1$ and $\Lambda \neq
0$ {}Friedmann-Lema\^{\i}tre models. I explicitly write the
second-order theories in terms of closed one-dimensional
integrals. In cosmologically interested cases ($\Lambda = 0$ or
$\Omega + \lambda = 1$), they reduce to elementary functions or
hypergeometric functions. For arbitrary $\Omega$ and $\Lambda$, I
present accurate fitting formula which are sufficient in practice for
the observational cosmology. It is reconfirmed for generic $\Omega$
and $\Lambda$ of interest that second-order effect only weakly depends
on these parameters.

\medskip

\bigskip

\begin{center}
{\it Progress of Theoretical Physics (Letters)}, in press
\end{center}

\end{titlepage}

\baselineskip=18pt
\setcounter{page}{2}

The gravitational instability presumably plays an essential role in
the formation of the large-scale structure of the universe. The
dynamics of such interesting phenomena involves the nonlinearity which
is difficult to deal with. Any exact solution for nonlinear evolution
is not known in general situation. We have been mainly resorted to
$N$-body methods for fully nonlinear problems in this field. Although
such methods can shed light on strongly nonlinear regime, the
resolution is fairly limited and they can barely survey only small
fraction of parameter space of possible models. On large-scales where
density fluctuations are small, perturbation theories are quite
powerful.  They can analytically give results for large parameter
space of possible models.

Two formulations for higher-order perturbation theories are focused in
the literature. One is Eulerian formulation
\cite{dor75,jus81,vis83,jus84,fry84,gor86,gri87,gri87b,col90,sut91,
mak92,ber92a,ber92b} and the other is Lagrangian formulation
\cite{mou91,buc92,bou92,gra93a,gra93b,buc93a,buc93b,lac93,buc94,cat95,
bou95,hiv95}. The first order in Eulerian formulation is well-known
linear theory and that in Lagrangian formulation corresponds to
well-known Zel'dovich approximation \cite{zel70,zel73}.

Since current observations seem to point to $\Omega < 1$ and/or
$\Lambda \neq 0$ \cite{pee91}, it is important to develop perturbation
theories in general $\Omega$ and $\Lambda$. Perturbation theory in
Eulerian space for Friedmann models (Throughout this letter, I
mean models with arbitrary $\Omega$ and $\Lambda = 0$ by Friedmann
models and models with arbitrary $\Omega$ and $\Lambda$ by
Friedmann-Lema\^{\i}tre models) are considered by several people
\cite{mar91,bou92,ber94a,ber95,cat94c}. Lagrangian-space counterpart
is considered by some authors
\cite{bou92,gra93a,gra93b,lac93,cat95,bou95}. The second-order
perturbation theories in Friedmann models can be expressed in closed
forms. For Friedmann-Lema\^{\i}tre models, however, the analytical
expression for the time-dependence in second-order perturbation
theories has not been known except for numerical solutions
\cite{ber94a,buc92,buc93a,buc94,bou95}.

In this {\it Letter}, I have obtained the explicit solution of the
time-dependence in second-order perturbation theories both in Eulerian
and Lagrangian space for the first time in models with arbitrary
$\Omega$ and $\Lambda$.

Let us briefly review second-order perturbation theories first. In
Friedmann-Lema\^{\i}tre models, non-relativistic pressure-less
self-gravitating fluid are governed by the following continuity
equation, Euler equation of motion and Poisson field equation
\cite{pee80}:
\begin{eqnarray}
   &&
   \dot{\delta} + \nabla\cdot[(1 + \delta)\bfv] = 0,
   \label{eq1}
   \\
   &&
   \dot{\bfv} + 2 H \bfv + (\bfv\cdot\nabla)\bfv +
   \nabla\Phi = \bf0,
   \label{eq2}
   \\
   &&
   \nabla^2 \Phi = \frac32 H^2 \Omega \delta,
   \label{eq3}
\end{eqnarray}
where $\bfx$, $\bfv(\bfx,t)$, $\Phi(\bfx,t)$ are position, peculiar
velocity, peculiar potential in comoving coordinate, respectively,
which correspond to $a\bfx$, $a\bfv(\bfx,t)$, $a^2\Phi(\bfx,t)$ in
physical units, respectively.  A dot denotes time derivative and
$\nabla \equiv \partial/\partial\bfx$ denotes spatial derivative with
respect to comoving coordinates.

In Eulerian perturbation theory, these dynamical equations are solved
perturbatively with respect to density contrast $\delta$ and velocity
$\bfv$, keeping only terms of leading time-dependence
\cite{pee80,fry84,gor86}. The perturbative expansion is denoted as
$\delta = \delta^{(1)} + \delta^{(2)} + \cdots$, $\bfv = \bfv^{(1)} +
\bfv^{(2)} + \cdots$ where $\delta^{(n)}$ and $\bfv^{(n)}$ are of
order $\left(\delta^{(1)}\right)^n$. The perturbative solution of the
velocity field derived from this procedure is irrotational, $\nabla
\times \bfv^{(n)} = \bf0$, in the leading time-dependence of
$\bfv^{(n)}$. Thus the perturbative velocity field is determined by
its divergence $\theta^{(n)} \equiv \nabla \cdot
\bfv^{(n)}/H$ as $\bfv^{(n)} = H \nabla\triangle^{-1} \theta^{(n)}$,
where $\triangle^{-1}$ denotes the inverse Laplacian. The result, up
to second-order \cite{pee80,ber94a,ber95}, is
\begin{eqnarray}
   &&
   \delta^{(1)}(\bfx,t) = D(t) \epsilon(\bfx),
   \label{eq4} \\
   &&
   \theta^{(1)}(\bfx,t) = - f(t) D(t) \epsilon(\bfx),
   \label{eq5} \\
   &&
   \delta^{(2)}(\bfx,t) =
   D^2(t)\left\{\frac12[1 + K(t)] \epsilon^2(\bfx) +
      \nabla\epsilon(\bfx) \cdot \nabla\varphi(\bfx)
      +
      \frac12[1 - K(t)]\varphi_{,ij}(\bfx) \varphi_{,ij}(\bfx) \right\},
   \label{eq6} \\
   &&
   \theta^{(2)}(\bfx,t) =
   - f(t) D^2(t)\left\{C(t)\epsilon^2(\bfx) +
      \nabla\epsilon(\bfx) \cdot \nabla\varphi(\bfx)
      +
      [1 - C(t)]\varphi_{,ij}(\bfx) \varphi_{,ij}(\bfx) \right\},
   \label{eq7}
\end{eqnarray}
where $\epsilon(\bfx)$ corresponds to the initial fluctuation and
$\varphi(\bfx)$ is its potential field, $\nabla^2 \varphi = \epsilon$
and $f \equiv \dot{D}/(HD)$, $K \equiv F/D^2$ $C \equiv
\dot{F}/(2D\dot{D})$ are defined using the growing solution $D$, $F$
of the following ordinary differential equations:
\begin{eqnarray}
   &&
   \ddot{D} + 2 H \dot{D} - \frac32 H^2 \Omega D = 0,
   \label{eq8} \\
   &&
   \ddot{F} + 2 H \dot{F} - \frac32 H^2 \Omega F =
   \frac32 H^2 \Omega D^2.
   \label{eq9}
\end{eqnarray}
The solution of the linear growth rate $D$ governed by Eq.~(\ref{eq8})
is well-known. In expanding universes, it is \cite{hea77,pee80}
\begin{equation}
   D(t) \propto a\Omega\int^1_0 dx
      [\Omega/x + \lambda x^2 + 1 - \Omega - \lambda]^{-3/2}.
   \label{eq10}
\end{equation}
Note that $\Omega = 8\pi G \rho/(3H^2)$, $\lambda = \Lambda/(3H^2)$
are time-dependent parameters. The function $f$ is similarly expressed
\cite{lah91}. The solution of Eq.~(\ref{eq9}) for Friedmann
models was explicitly given \cite{bou92}. Although in $\Omega +
\lambda = 1$ models, this equation have been investigated numerically
\cite{ber94a,bou95}, I will give the closed representation of the
solution of this equation for arbitrary $\Omega$ and $\lambda$
universes below.

Lagrangian perturbation theory considers motion of mass elements
labeled by unperturbed Lagrangian coordinates $\bfq$. The comoving
Eulerian position of mass element $\bfq$ at time $t$ is denoted by
$\bfx(\bfq,t)$. The displacement field $\bfPsi(\bfq,t)$ defined by
\begin{equation}
   \bfx(\bfq,t) = \bfq + \bfPsi(\bfq,t),
   \label{eq11}
\end{equation}
is a dynamical variable in this formulation. Density contrast and a
velocity field are derived from a displacement field as
\begin{eqnarray}
   &&
   \delta[\bfx(\bfq,t),t] = J^{-1} - 1,
   \label{eq12} \\
   &&
   \bfv[\bfx(\bfq,t),t] = \dot{\bfx}(\bfq,t) = \dot{\bfPsi}(\bfq,t),
   \label{eq13}
\end{eqnarray}
where $J \equiv \det[\partial x_i/\partial q_j] = \det[\delta_{ij} +
\partial \Psi_i/\partial q_j]$ is the Jacobian from Lagrangian space
to Eulerian space. A dot denotes time derivative fixing Lagrangian
coordinates $\bfq$, so should not be confused with the Eulerian time
derivative in Eq.~(\ref{eq1})-(\ref{eq3}). These Eulerian
Eqs.~(\ref{eq1})-(\ref{eq3}) are transformed to the Lagrangian
equations governing displacement field as
\begin{eqnarray}
   &&
   J \nabla_x \cdot [\ddot{\bfPsi} + 2H \dot{\bfPsi}]
   + \frac32 H^2 \Omega (1 - J) = 0,
   \label{eq14} \\
   &&
   \nabla_x \times [\ddot{\bfPsi} + 2H \dot{\bfPsi}] = \bf0.
   \label{eq15}
\end{eqnarray}
Again, a dot denotes Lagrangian time derivative and $\nabla_x$ is the
spacial derivative with respect to Eulerian coordinates $\bfx$, so
$(\nabla_x)_i = J^{-1} \widetilde{J}_{ij}(\nabla_q)_j$, where
$\widetilde{J}_{ij}$ are cofactors of the Jacobian $J$.

In usual treatment of Lagrangian perturbation theory, one assumes an
additional condition, i.e., vorticity-free condition:
\begin{equation}
   \nabla_x \times \bfv = \nabla_x \times \dot{\bfPsi} = \bf0.
   \label{eq16}
\end{equation}
This is not unreasonable requirement because vorticity is expected to
be diluted by expansion before `turn-around' in some sense
\cite{pee80}. The condition (\ref{eq16}) is a sufficient condition for
dynamical equation (\ref{eq15}) \cite{buc92}, so the solutions with
vorticity-free condition are in a subclass of the general solutions
(rotational perturbation is argued in the literature
\cite{buc92,buc93a}). This subclass is achieved if one limit the
initial condition to irrotational field because irrotational velocity
field at one time remains irrotational at later times from Kelvin's
circulation theorem.  Therefore, Eqs.~(\ref{eq14}) and (\ref{eq16})
are solved perturbatively for displacement field $\bfPsi =
\bfPsi^{(1)} + \bfPsi^{(2)} + \cdots$, keeping only terms of leading
time-dependence. For irrotational initial displacement field, the
result, up to second-order (see \cite{buc93a,bou95} for detail), is
\begin{eqnarray}
   &&
   \bfPsi^{(1)} = - D(t) \nabla \triangle^{-1} \epsilon(\bfx),
   \label{17} \\
   &&
   \bfPsi^{(2)} = - \frac12 K(t) \nabla \triangle^{-1}
   \sum_{i\neq j}\left( \Psi^{(1)}_{i,i} \Psi^{(1)}_{j,j} -
   \Psi^{(1)}_{i,j} \Psi^{(1)}_{i,j}\right).
   \label{eq18}
\end{eqnarray}
The first-order solution is equivalent to the Zel'dovich approximation
\cite{zel70,zel73}. The factors $D$ and $K$ are the same quantities in
Eulerian perturbation theory [Eqs.~(\ref{eq4})-(\ref{eq7})]. Thus the
importance of solving differential Eq.~(\ref{eq9}) in a closed
form applies to second-order Lagrangian perturbation theory as well.

The explicit form of the growing solution of differential equation
(\ref{eq9}) is now presented below. Bernardeau \cite{ber94a} realized
that the solution of spherical collapse is useful in this
problem. Although he claimed to obtain explicit results only for
Friedmann models, I will show that this approach can lead to the
explicit closed solution even in Friedmann-Lema\^{\i}tre models.

When the density contrast and the velocity field are spherically
symmetric, Eqs.~(\ref{eq1})-(\ref{eq3}) or Eqs.~(\ref{eq14}) and
(\ref{eq15}) imply that the physical radius $r$ of each mass-shell
obeys the following equation:
\begin{equation}
   \ddot{r} = - \frac{GM}{r^2} + \frac{\Lambda}{3}\; r,
   \label{eq19}
\end{equation}
where $M$ is the total mass inside the mass-shell. This happens to be
the same as Tolman-Bondi equation which describe the motion of
mass-shell of spherically symmetric dust in general relativity, in
spite of our Newtonian treatment. The density contrast at the center
is evaluated by $r \rightarrow 0$ limit of the following quantity:
\begin{equation}
   \Delta(t) = \frac{2GM}{H^2 \Omega r^3} - 1,
   \label{eq20}
\end{equation}
which satisfies
\begin{equation}
   \ddot{\Delta} + 2H\dot{\Delta} - \frac32 H^2 \Omega \Delta =
   \frac32 H^2 \Delta^2 + \frac43 \frac{\dot{\Delta}^2}{1 + \Delta}.
   \label{eq21}
\end{equation}

I consider the solution $\Delta(t;\delta_{\rm i})$ with the initial
condition, $\Delta(t_{\rm i}) = \delta_{\rm i}$ and
$\dot{\Delta}(t_{\rm i}) = 0$, which is expanded in Taylor series with
respect to $\delta_{\rm i}$:
\begin{equation}
   \Delta(t;\delta_{\rm i}) =
   D(t) \delta_{\rm i} +
   \frac12 D_2(t) \delta_{\rm i}^{\,2} + \cdots.
   \label{eq22}
\end{equation}
The coefficient $D$ obeys the differential equation (\ref{eq8}) of
linear growth rate and $F = 3D_2/2 - 2D^2 $ obeys the differential
equation (\ref{eq9}). Thus, all we have to do is integrating
Eq.~(\ref{eq19}) and then evaluating the coefficient of the Taylor
series (\ref{eq22}). The specific choice of the initial condition,
$\dot{\Delta}(t_{\rm i}) = 0$, does not lose the generality because we
are interested in a growing mode.

Integrating Eq.~(\ref{eq19}), we have
\begin{equation}
   t - t_{\rm i} =
   \int_{r_{\rm i}}^r dr
   \left(\frac{2GM}{r} + \frac{\Lambda}{3} r^2 + 2 E\right)^{-1/2},
   \label{eq23-1}
\end{equation}
where a suffix i indicates initial values. $E$ is an integration
constant,
\begin{eqnarray}
   2 E &=& \dot{r_{\rm i}}^{\,2} - \frac{2GM}{r_{\rm i}} -
   \frac{\Lambda}{3} {r_{\rm i}}^2
   \nonumber \\
   &=& \left(\frac{2GMH}{\Omega}\right)^{2/3}
   \left(
      1 - \Omega - \lambda - \frac{a}{a_{\rm i}}\Omega\delta_{\rm i}
   \right)
   (1 + \delta_{\rm i})^{-2/3}.
   \label{eq23-3}
\end{eqnarray}
In the second line of the above equations, following equations are
used:
\begin{eqnarray}
   &&
   \dot{r_{\rm i}} = H_{\rm i} r_{\rm i}
   \nonumber\\
   &&
   r_{\rm i} =
   \left(\frac{2GM}{{H_{\rm i}}^2\Omega_{\rm i}}\right)^{1/3}
   (1 + \delta_{\rm i})^{-1/3} =
   \frac{a_{\rm i}}{a}
   \left(\frac{2GM}{H^2 \Omega}\right)^{1/3}
   (1 + \delta_{\rm i})^{-1/3}.
   \nonumber
\end{eqnarray}
Using a new integration variable $x = (H^2\Omega/2GM)^{1/3}r$,
Eq.~(\ref{eq23-1}) reduces to
\begin{eqnarray}
   &&
   H(t - t_{\rm i}) =
   \int_{(1 + \delta_{\rm i})^{-1/3}
   a_{\rm i}/a}^{(1 + \Delta)^{-1/3}}
   dx
   \left[\frac{\Omega}{x} + \lambda x^2
    +
    \left(1 - \Omega - \lambda -
       \frac{a}{a_{\rm i}}\Omega \delta_{\rm i} \right)
      (1 + \delta_{\rm i})^{-2/3}\right]^{-1/2}.
   \label{eq23}
\end{eqnarray}
Partially differentiating this equation with respect to $\delta_{\rm
i}$, the partial derivative $\partial\Delta/\partial\delta_{\rm i}$ is
obtained as follows:
\begin{eqnarray}
   &&
   \frac{\partial\Delta}{\partial \delta_{\rm i}} =
   (1 + \Delta)^{3/2}
   \sqrt{\Omega + \lambda(1 + \Delta)^{-1} + A(1 + \Delta)^{-1/3}}
   \nonumber \\
   &&
   \qquad\qquad\qquad
   \times\;\left[
   B - \frac32\frac{\partial A}{\partial \delta_{\rm i}}
   \int_{(1 + \delta_{\rm i})^{-1/3}a_{\rm i}/a}^{(1 + \Delta)^{-1/3}}
   dx
   \left(\frac{\Omega}{x} + \lambda x^2 + A\right)^{-3/2}\right],
   \label{eq23-4}
\end{eqnarray}
where
\begin{eqnarray}
   &&
   A = (1 - \Omega - \lambda - \Omega \delta_{\rm i}a/a_{\rm i})
   (1 + \delta_{\rm i})^{-2/3}
   \nonumber \\
   &&
   B = \frac{(1 + \delta_{\rm i})^{-3/2} (a_{\rm i}/a)^{3/2}}
       {\sqrt{\Omega + \lambda(1 + \delta_{\rm i})^{-1}
        (a_{\rm i}/a)^3 + A (1 + \delta_{\rm i})^{-1/3}a_{\rm i}/a}}.
   \nonumber
\end{eqnarray}
Partially differentiating Eq.~(\ref{eq23-4}) again,
$\partial^2\Delta/\partial\delta_{\rm i}^{\,2}$ is obtained similarly,
though it is lengthy. Substituting $\delta_{\rm i} = 0$ for these
expressions, $D$ and $D_2$ are obtained explicitly. Note that
$\delta_{\rm i} = 0$ implies $\Delta = 0$ because of vanishing
perturbation.

In expanding universes, the term of leading time-dependence
corresponds to the leading term of $a/a_{\rm i}$ of the resulting
expression. With this prescription, one can reproduce Eq.~(\ref{eq10})
for $D$ as expected. Similarly, I find the explicit expression for
$D_2$ and therefore for $F$. The final results are relatively simple:
\begin{eqnarray}
   &&
   K(\Omega,\lambda) =
   \frac{\Omega}{4} - \frac{\lambda}{2} -
   \frac{1}{U_{3/2}} \left[ 1 - \frac32
   \frac{U_{5/2}}{U_{3/2}}\right],
   \label{eq24} \\
   &&
   C(\Omega,\lambda) =
   \frac{1}{8f} \left\{ -3\Omega \rule{0mm}{3.5ex}
   +
   \frac{1}{(U_{3/2})^2}
   \left[2 + 4 U_{3/2} - 3(2 + \Omega - 2\lambda) U_{5/2}\right]
   \right\},
   \label{eq25}
\end{eqnarray}
where
\begin{equation}
   U_\alpha(\Omega,\lambda) =
   \int_0^1 dx
   [\Omega/x + \lambda x^2 + 1 - \Omega - \lambda]^{-\alpha}.
   \label{eq26}
\end{equation}
These are the main results of this {\it Letter}. In Einstein-de
Sitter, Friedmann and flat models, $U_\alpha(1,0) = 1/(1 + \alpha)$,
$U_\alpha(\Omega,0) = F(1,\alpha,\alpha + 2;1 - \Omega)/(1 + \alpha)$,
and $U_\alpha(\Omega,1 - \Omega) = F(1,\alpha,(\alpha + 4)/3;1 -
\Omega)/(1 + \alpha)$, where $F$ is the hypergeometric
function. Moreover, in Friedmann models, $U_{3/2}$ and $U_{5/2}$ are
are actually elementary functions which reproduce the previous results
\cite{bou92}.

For practical purposes, Eqs.~(\ref{eq24}) and (\ref{eq25}) are
accurately approximated by
\begin{eqnarray}
   &&
   K \approx \frac37 \Omega^{-1/30} -
   \frac{\lambda}{80}
   \left(1 - \frac32 \lambda \log_{10}\Omega \right),
   \label{eq27} \\
   &&
   C \approx \frac37 \Omega^{-11/200} -
   \frac{\lambda}{70}
   \left(1 - \frac73 \lambda \log_{10}\Omega \right),
   \label{eq28}
\end{eqnarray}
within maximum error 0.6\% for both for $-1 \leq \log_{10}\Omega \leq
0$ and $0 \leq \lambda \leq 1$ . Quite obviously from this fitting,
dependence on parameters $\Omega$ and $\lambda$ is weak. In fact $K$
and $C$ for the above ranges of $\Omega$ and $\lambda$ are different
from the Einstein-de~Sitter values within 8\% and 14\%, respectively.
Thus second-order effect is practically very insensitive either to
$\Omega$ or to $\lambda$. Since this level of errors is negligible
compared with the uncertainties with the current observational data,
comparison of data and theories in a weakly nonlinear regime provides
the test of gravitational instability almost independently of the
poorly determined $\Omega$ and $\lambda$. Skewness
\cite{gor86,bou92,jus93,luo93,fry94b,ber94a,lok95,cat94c}, probability
distribution function \cite{jus95,ber94b,ber95b} and topology
\cite{mat94,mat95} in a weakly nonlinear regime are among such
directly observable quantities.

The analysis of weakly non-linear regime will be important as future
redshift surveys will be available and statistical ambiguity on large
scales will decrease.

\vskip1.0cm

I thank Y.~Suto for a careful reading of the manuscript and useful
comments, and T.~T.~Nakamura for stimulating discussions.  I
acknowledges the fellowship from the Japan Society of Promotion of
Science.  This research was supported in part by the Grants-in-Aid by
the Ministry of Education, Science and Culture of Japan (No.~0042)

\end{document}